# DYNAMICS OF CHEMICAL REACTIONS: OPEN EQUILIBRIUM

*Do we need coefficients of thermodynamic activity?*

B. Zilbergleyt
Independent Scholar
E-mail: LIVENT1@MSN.COM

ABSTRACT.

Dynamic properties of chemical reactions and appropriate relationships for open chemical equilibrium are discussed in approach of the chemical dynamics. New way to calculate composition of chemical systems in open equilibrium, based on the amended equation for the change of Gibbs' free energy, was exemplified using reactions of cobalt double oxides with sulfur. Evaluated this way equilibrium reaction extents had a good match with values found by direct thermodynamic simulation. The developed method allows us to perform analysis of complex chemical systems without coefficients of thermodynamic activity. In cases when the coefficients are in need or just desirable, the method offers a simple way to calculate them. Possible systems to apply method of chemical dynamics are discussed.

Previous paper [1] has described new approach to the thermodynamics of open chemical systems. With assigned generic name "chemical dynamics", that approach is based on assumption that there is a linear relationship between external thermodynamic force and relevant shift of chemical reaction extent in the vicinity of "true" equilibrium. Corresponding expression for reduced Gibbs' free energy change may be written as

$$\Delta g_i = \Delta g^0_i + f_{ie}^* \Delta_i + \ln \Pi a_{ki}(\Delta_i, \eta_{ki}) + \tau_i \Delta_i \delta_i, \qquad (1)$$

where $\Delta_i$ - symbol for reaction extent $\Delta\xi$, in "true" equilibrium $\Delta_i = 1$; $\delta_i$ – symbol for $\delta\xi = (1-\Delta)$ - reaction shift from equilibrium; $\Delta g_i = \Delta G_i/RT$ – reduced change of the reaction Gibbs' free energy; $f_{ie}^* \Delta_i$ – reduced (divided by RT) external thermodynamic force multiplied by $\Delta$ thus having dimension of energy; $a_{ki}(\Delta_i, \eta_{ki})$ - thermodynamic activity; $\eta_{ki}$ – amount of moles of k-participant of i-reaction, consumed/arrived in the run of this reaction from initial state to "true" thermodynamic equilibrium; and $\tau$ is a parameter of the theory characterizing system interaction with its environment. We have introduced the $\tau_i \Delta_i \delta_i$ product in [1] as "chaotic" member due to analogy with the well known chaotic equation [2].
Taking into account that $\Delta g^0_i = -\ln K_i$, or $\Delta g^0_i = \ln \Pi`_i$, tick mark and asterisk relate values to isolated (ISEQ) and open equilibrium (OPEQ) correspondingly, equation (1) will turn to

$$\Delta g_i = f_{ie}^* \Delta_i - \ln(\Pi`_i/\Pi^*_i) + \tau_i \Delta_i \delta_I. \qquad (2)$$

One of the most important results described in [1] was a specific Hooke's law type of shift-force dependency, which is linear up to a certain value of the force, and then shows a kind of saturation. In many cases the curve is still linear (more correctly, quasi-linear) beyond the bend point but under different slope.

In the run of this work about a hundred of reactions with essential negative $\Delta G^0$ values were investigated, and in no cases deviations from the chemical Hookes law were found. Typical distribution of reaction coordinates in open equilibrium vs. thermodynamic forces for a large group of reactions ($Me_1O+S$) at 298K in systems, where primary oxide is bound into double oxide $Me_1O*Me_2O$ (MeO*R), is shown on Fig.1. Related source data is in the Table1. Zero thermodynamic force corresponds to conditional interaction of an oxide with itself thus representing isolated system, and $\Delta^*_i = 1$ and $\delta^*_i = 0$.

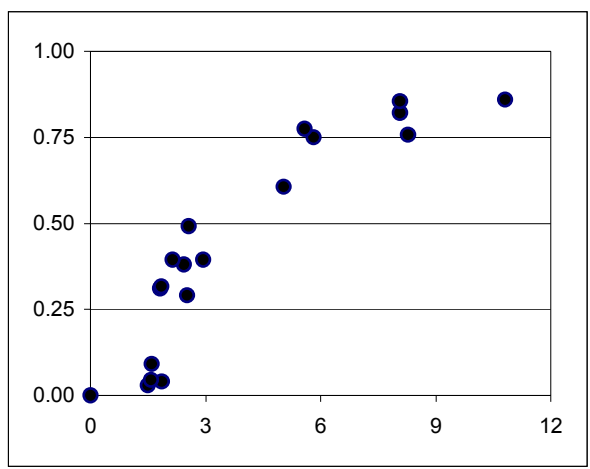

Fig.1. Correlation between reaction shift $\delta^*$ (ordinate) and external thermodynamic force (abscissa) at open equilibrium in a group of various metal double oxide reactions with sulfur.

Table 1.
Thermodynamic force and reaction shift in MeO*R-S systems.

| Double oxides | $(\Delta g^0)/\Delta\xi$, kJ/m | $\delta$ |
|---|---|---|
| MeO∗MeO | 0,00 | 0,00 |
| FeO *SiO2 | 1,50 | 0,03 |
| FeO *Cr2O3 | 1,86 | 0,04 |
| NiO *SiO2 | 1,60 | 0,09 |
| NiO *Cr2O3 | 2,51 | 0,29 |
| CoO *Fe2O3 | 1,82 | 0,31 |
| FeO *Fe2O3 | 1,84 | 0,32 |
| CoO *TiO2 | 2,44 | 0,38 |
| CoO *WO3 | 2,14 | 0,39 |
| CoO *Cr2O4 | 2,94 | 0,39 |
| NiO *WO3 | 2,55 | 0,49 |
| FeO *Al2O3 | 5.03 | 0,61 |
| MnO*SiO2 | 5,81 | 0,75 |
| PbO *B2O3 | 8,28 | 0,76 |
| MnO*Fe2O3 | 5,58 | 0,78 |
| PbO *SiO2 | 8,07 | 0,82 |
| PbO *WO3 | 8,07 | 0,85 |
| PbO *TiO2 | 10,81 | 0,86 |

Physical nature of the parameter $\tau$ wasn't clear at the time when [1] was published. Further investigation has unambiguously showed that this parameter has a unique value for every chemical reaction given the stoichiometric equation and reactant amounts, and is responsible for reaction resistance against external thermodynamic force. It makes very clear why combination of $\tau$ and external thermodynamic force defines the reaction shift from "true" equilibrium.

Each chemical reaction can be characterized by several dynamic curves. They are already familiar force-shift curve and two others - reaction extent vs. effective Gibbs' free energy change, calculated as a sum of internal and external values, and reaction extent vs. "chaotic" member $\tau_i\Delta_i\delta_i$. Shift-force curve that can be used to determine $\tau$ for the simplest reaction A+B=AB is shown on Fig.2. Next picture, Fig.3 shows a joint graph of above mentioned curves. An arbitrary value of 0.5 was assumed for $\eta_A$, thus unilaterally defining $\Delta g^0$ for that reaction. Data used for plotting the graphs are in the Table 2.

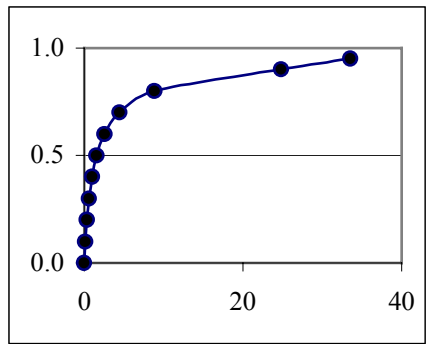

Fig.2. Shift-force curve for reaction A+B=AB with $\eta_A$= 0.5.

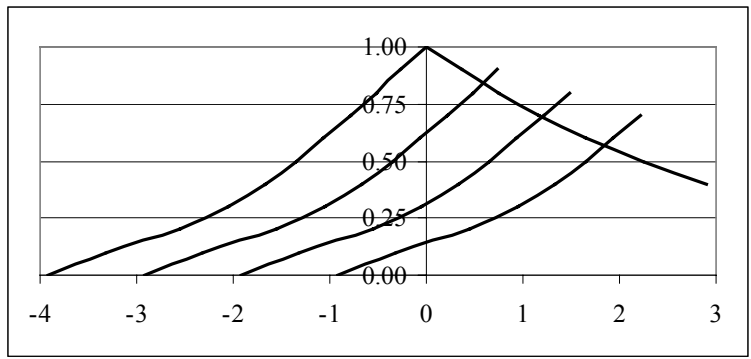

Fig.3. Reaction extent $\Delta$ (ordinate) vs. (1) $[-\ln(\Pi`/\Pi(\Delta))+f_{ie}*\Delta_i]$ (ascending curves), and vs. (2) "chaotic" term $\tau_i\Delta_i\delta_I$ (the descending curve). Points of intersections of the descending curve with all ascending curves define appropriate reaction extents at ISEQ ($f_{ie}*\Delta_i = 0$, $\Delta_i = 1$, the leftmost), and OPEQ ($f_{ie}*\Delta_I = \Delta G^0_{CoO*R}$, $\Delta_i < 1$, we have taken arbitrary real numbers of 1, 2, 3 for it). Reaction A+B=AB with $\eta_A$=0.5.

Quotient of the activity products (columns 3 and 4) was calculated as

$$\Pi`/\Pi(\Delta_i)=(1/\Delta_i)*[(2-\eta)/(2-\Delta_i*\eta)]*[(1-\Delta_i*\eta)/(1-\eta)]^2. \qquad (2)$$

Now we will give one practical example. Consider reaction between cobalt double oxide and sulfur at initial amounts of reactants $n^0$ (CoO*R) = 1m and $n^0$ (S) = 2m, where *R represents one of the non-reacting with sulfur oxides $TiO_2$, $Cr_2O_3$, $WO_3$

$$2CoO*R + 4S = CoS_2 + CoS + SO_2 + 2R. \qquad (3)$$

Table 2.

Data for plotting Fig2 and 3.

| $\Delta$ | $\delta$ | $\ln(\Pi`/\Pi(\Delta_i))$ | $[-\ln(\Pi/\Pi(\Delta_i))/\Delta_i]$ | $\tau_i$ | $\tau_i \Delta_i \delta_i$ |
|---|---|---|---|---|---|
| 0.00 | 1.00 | -3.91 | 44.64 | | |
| 0.10 | 0.90 | -3.32 | 33.24 | | |
| 0.20 | 0.80 | -2.55 | 12.74 | | |
| 0.30 | 0.70 | -2.06 | 6.85 | 10.25 | 2.15 |
| 0.40 | 0.60 | -1.67 | 4.18 | 7.18 | 1.72 |
| 0.50 | 0.50 | -1.35 | 2.70 | 5.51 | 1.38 |
| 0.60 | 0.40 | -1.06 | 1.76 | 4.47 | 1.07 |
| 0.70 | 0.30 | -0.79 | 1.12 | 3.77 | 0.79 |
| 0.80 | 0.20 | -0.52 | 0.65 | 3.28 | 0.53 |
| 0.90 | 0.10 | -0.26 | 0.29 | 2.92 | 0.26 |
| 1.00 | 0.00 | 0.00 | 0.00 | 2.33 | 0.00 |

Thermodynamic equivalent of transformation $\eta_{CoO*R} = 0.9073$ at 1000K was found using HSC Chemistry thermodynamic simulation of reaction

$$2CoO + 4S + 2Y_2O_3 = CoS_2 + CoS + SO_2 + 2Y_2O_3. \qquad (3)$$

with neutral (non-reacting with sulfur at chosen temperature) diluent $Y_2O_3$, ratio $CoO:Y_2O_3 = 1$. Shift-force curve for this reaction is shown on Fig.4.

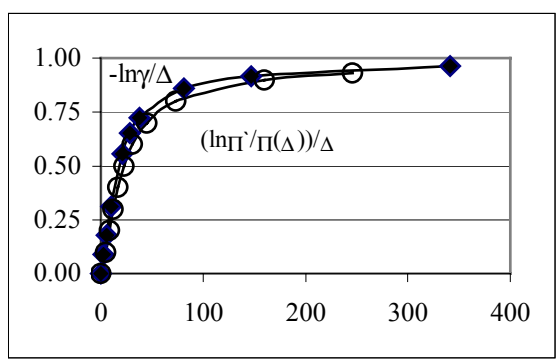

Fig.4. Shift-force graphs for reaction (3).

It was plotted with data calculated for the force as $(-\ln \gamma_{CoO})/\Delta$ and $(-\ln(\Pi`/\Pi(\Delta)))/\Delta$. The match between both on Fig.4 is remarkable. Values of the parameter $\tau$ for $0 =< \delta => 0.40$ found from the graph in the first case was 32.61, in the second 41.27. Joint graph for this reaction with data based on $\tau = 41.27$ is plotted on Fig.4. Points of intersection of ascending curves with the descending curve, corresponding to appropriate open equilibria, are shown on an enlarged part of Fig.6. Comparison of simulated with HSC values of reaction extents with those estimated from Fig.5 is given in Table 3.

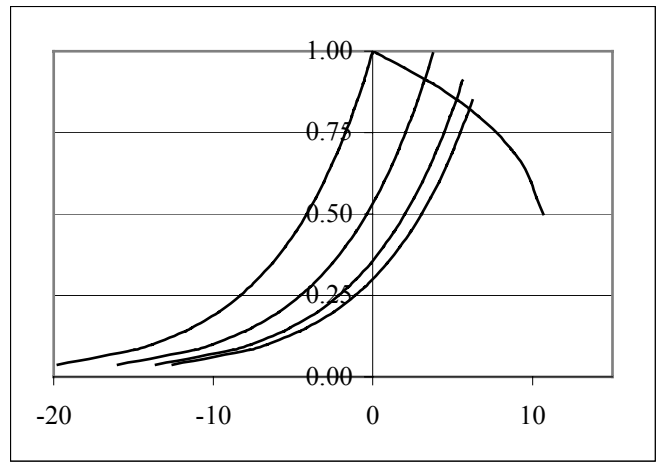

Fig.5. Reaction extent Δ vs. reduced effective Gibbs' free energy changes and vs. "chaotic" term $\tau\Delta\delta$ (the descending curve). Ascending curves correspond to, left to right, CoO and CoO*TiO$_2$, *Cr$_2$O$_3$, *WO$_3$. Once again, the leftmost curve meets "chaotic" term at Δ=1. Reaction $2CoO+4S+2Y_2O_3 = CoS2+CoS+ SO2+2Y_2O_3$.

Table 3.

Simulated and graphically evaluated equilibrium values or reaction extents at OPEQ. Values of $\Delta G^0_{CoO*R}$ were taken from HSC Chemistry database.

|  | CoO*TiO$_2$ | CoO*Cr$_2$O$_3$ | CoO*WO$_3$ |
|---|---|---|---|
| $(-\Delta G^0_{CoO*R}/RT)$ | 3.77 | 6.17 | 7.20 |
| Δ sim., HSC | 0.92 | 0.89 | 0.85 |
| Δ est., τ = 41.27 | 0.92 | 0.85 | 0.82 |
| Δ est., τ = 32.61 | 0.90 | 0.82 | 0.77 |

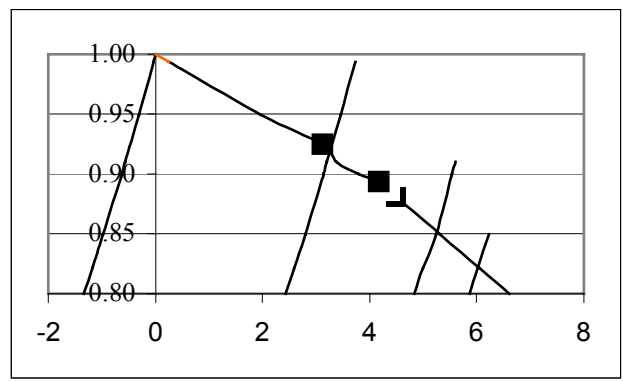

Fig.6. Rescaled graph of Δ (ordinate) vs. "chaotic" member (abscissa). Simulated with HSC values are shown as points on the descending curve. They follow in the same order as the ascending curves for used double oxides.

In the table, values Δ sim. and Δ est. were determined using two in principle different methods, the last of them splits in its turn by two also slightly different. For graphic evaluation in a system, taken arbitrarily out of the model set, the match is good and proves ability of the method of chemical dynamics to simulate open equilibria and complex chemical systems correctly. Better match for the values found purely by our method is a pleasant but yet not sufficient for far going conclusions and excessive optimism fact. So far we are investigating the method in general.

Results of this work show a possibility to find equilibrium composition of a subsystem without simulation the complex chemical system as a whole. We can do it if we know how to write down external thermodynamic force, through which the "complimentary" part of the system is acting against the subsystem. Along with results of [1] they prove that with the method of chemical dynamics one can avoid usage of thermodynamic activity coefficients at all. Nevertheless, in cases when the coefficients are still in need or just more habitual, the method of chemical dynamics provides quite simple way to find them using shift-force relationship. This relationship has been found to be unilateral within the entire possible scope of reaction shifts (from 0 up to almost 1) thus expanding the opportunities out of linear (more exactly, quasi-linear) area.

In many cases this can eliminate expensive experimental works to define coefficients of thermodynamic activity.

Double oxides with strong bounds between components, used in this work as a model set, represent only one example out of a big number of possible applications. Next and probably more practically interesting is application to the systems with reactions having comparable energetic characteristics (supposed next step of this work). Besides that, one interesting application of the method of chemical dynamics might also occur in stationary systems. In this case, using Onzager's relation between flows and thermodynamic forces, one can find equivalent value of external force to consider the stationary state formally related to the open equilibrium with chemical reaction shifted by this force. Most probably, this statement can be easily checked out in electrochemical systems, where electrochemical forces are definitely external to chemical reactions in the electrochemical cell, and relationships between them and flows (that is, electrical currents) are well known.

To conclude, it is worthy to mention that ideology of here developed method is totally different from classical way of analysis of complex systems. Conventional method aggregates all parts of the system to analyze the whole on a probability basis. We go opposite direction, dividing the system by parts and replacing probability treatment of subsystems interaction by shift-force relationships. Difference between two approaches was in general well discussed in [3]. Developed in this work approach descends back to d'Alembert principle of classical mechanics [4, 5].